# Virtual Reality for Urban Walkability Assessment

Viet Hung Pham *
Eindhoven University of Technology
Eindhoven, The Netherlands
RMIT University
Melbourne, Australia
v.h.pham@tue.nl

Malte Wagenfeld
RMIT University
Melbourne, Australia
malte.wagenf eld@rmit.edu.au

Regina Bernhaupt
Eindhoven University of Technology
Eindhoven, The Netherlands
r.bernhaupt@tue.nl

## Abstract
Traditional urban planning methodologies often fail to capture the complexity of contemporary urbanization and environmental sustainability challenges. This study investigates the integration of Generative Design, Virtual Reality (VR), and Digital Twins (DT) to enhance walkability in urban planning. VR provides distinct benefits over conventional approaches, including 2D maps, static renderings, and physical models, by allowing stakeholders to engage with urban designs more intuitively, identify walkability challenges, and suggest iterative improvements. Preliminary findings from structured interviews with Eindhoven residents provide critical insights into pedestrian preferences and walkability considerations. The next phase of the study involves the development of VR-DT integrated prototypes to simulate urban environments, assess walkability, and explore the role of Generative Design in generating adaptive urban planning solutions. The objective is to develop a decision-support tool that enables urban planners to incorporate diverse stakeholder perspectives, optimize pedestrian-oriented urban design, and advance regenerative development principles. By leveraging these emerging technologies, this research contributes to the evolution of data-driven, participatory urban planning frameworks aimed at fostering sustainable and walkable cities.

## 1 Introduction
Walkability is a key consideration in urban planning, shaped by infrastructure, land use, and socio-economic factors. As cities face increasing urbanization and sustainability challenges, promoting walkable environments has become essential for reducing car dependency and fostering healthier, more sustainable communities [6, 15]. Research links walkability to public health benefits, including increased physical activity and improved mental well-being [7]. However, traditional urban planning methods often rely on quantitative metrics, overlooking subjective aspects such as safety, comfort, and aesthetics [3, 16]. This gap underscores the need for human-centric approaches that integrate pedestrian experiences into planning. Virtual Reality (VR) offers a solution by enabling immersive visualization and stakeholder engagement, allowing planners to assess walkability beyond static models. Integrating VR into urban design can enhance decision-making, bridging the gap between objective planning metrics and the experiential quality of walkable environments.

This research integrates VR and Digital Twin technology to provide planners and stakeholders with immersive, real-time simulations of urban environments, allowing for interactive evaluation and modification of design proposals. VR enhances spatial understanding and facilitates informed feedback, fostering collaborative decision-making. Digital Twin technology complements this by creating dynamic, data-driven urban models that reflect real-world conditions. Additionally, generative design algorithms are employed to explore a range of design possibilities, generating optimal solutions based on predefined walkability criteria, constraints, and urban planning objectives.

## 2 Research Gap
Urban walkability assessment remains a critical yet underdeveloped aspect of contemporary urban planning. While traditional methods rely on quantitative metrics such as pedestrian density, network connectivity, and land use, they often fail to capture the nuanced experiential and behavioral factors influencing walkability [3, 16]. The increasing adoption of data-driven urban planning tools, such as Geographic Information Systems (GIS), Big Data analytics, and Digital Twins, has enhanced analytical capabilities but remains limited in providing immersive, first-person assessments of pedestrian environments [2, 14]. This gap necessitates exploring VR as an interactive and experiential medium for walkability evaluation.

Current walkability assessments prioritize objective measurements like street connectivity and infrastructure availability but overlook subjective factors such as perceived safety, comfort, and environmental aesthetics [6, 15]. These traditional models fail to incorporate real-time human interaction with urban spaces, limiting their ability to simulate pedestrian behavior dynamically [7]. Without accounting for the experiential aspects of urban mobility, planning decisions may not align with the current growth of the urban environment, reducing the effectiveness of pedestrian-centered urban design [12].

Urban planners and designers primarily rely on static tools such as maps, zoning plans, and 3D renderings to communicate urban design proposals. While these tools provide useful overviews, they lack real-world dynamism, making it difficult to assess how design choices impact pedestrian movement, wayfinding, and spatial perception [8, 9]. Conventional simulations struggle to replicate varying environmental conditions, such as fluctuating traffic patterns, weather changes, and crowd dynamics, which are crucial for evaluating walkability comprehensively [18]. The inability to integrate real-time pedestrian interactions and scenario-based testing leaves gaps in understanding the full impact of urban design interventions [19].

Effective urban planning requires input from multiple stakeholders, including policymakers, urban planners, designers, and local communities. However, traditional planning processes often create barriers to public engagement due to the technical complexity of interpreting urban design proposals [11]. Many community members lack the expertise to analyze static maps or architectural blueprints, limiting their ability to provide meaningful feedback [17]. The lack of accessible visualization tools contributes to a disconnect between planners and the public, reducing the inclusivity and participatory nature of urban development processes.

Among emerging technologies, VR has demonstrated the potential to enhance urban planning by offering immersive and interactive urban simulations [1, 13]. Unlike static visualization tools, VR allows users to explore and experience pedestrian pathways, evaluate accessibility, and test real-time interactions within a dynamic urban simulation. This technology enhances comparative analysis by enabling planners to toggle between different design scenarios instantly, a capability particularly valuable for evaluating factors such as street width, placement of green spaces, and the visibility of key landmarks—elements that significantly influence pedestrian behavior but are difficult to assess through conventional methods [20, 10]. Additionally, integrating VR with DT technologies enables continuous urban monitoring, allowing for adaptive urban planning based on real-time data [4, 5].

Addressing these research gaps will contribute to developing more responsive, data-informed, and pedestrian-friendly urban environments. Future studies should focus on refining VR's capabilities to support large-scale urban assessments, incorporating quantitative analytics for objective evaluation, and enhancing accessibility to ensure broader adoption in urban planning workflows. By integrating immersive technologies with participatory planning, urban development can better reflect both the functional and experiential needs of pedestrians.

## 3 Transforming Walkability Assessment and Urban Planning Through VR

Traditional urban planning tools rely heavily on static visualizations such as maps, blueprints, and GIS-based models. While these tools provide valuable data on infrastructure and accessibility, they fall short of capturing how pedestrians experience urban spaces in real-time. VR addresses these limitations by offering an interactive, immersive environment that enables planners and stakeholders to assess walkability from a human-centered perspective. VR enhances spatial understanding, allows for scenario-based simulations, and fosters more effective communication among urban stakeholders.

### 3.1 First-Person Spatial Experience

VR provides an immersive, first-person perspective that allows users to navigate urban environments dynamically. Unlike 2D maps or static 3D renderings, VR offers real-time interaction with urban features, helping planners evaluate pedestrian pathways, crossings, and accessibility with greater accuracy. By simulating real-world movement, VR enables users to experience urban spaces as they would in everyday life, uncovering obstacles, inefficiencies, and areas for improvement that may not be evident in traditional design processes.

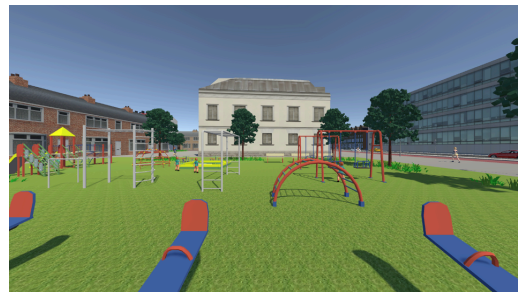

Figure 1: First person view in VR urban walkability assessment prototype.

One of VR's key advantages is its ability to provide a sense of scale and spatial relationships that influence walkability. Users can perceive the width of sidewalks, the proximity of green spaces, and the positioning of wayfinding signage in a way that static models cannot replicate. This enables urban planners to refine designs based on how people experience and move through the built environment. By integrating VR into walkability assessments, planners can identify and mitigate potential design flaws before construction, reducing costly modifications and enhancing pedestrian comfort.

### 3.2 Scenario Simulation

Urban environments are dynamic and shaped by changing conditions such as lighting, weather, traffic density, and pedestrian flow. Traditional urban planning tools struggle to account for these factors in real-time, making it difficult to predict how walkability changes under different scenarios. VR addresses this challenge by allowing planners to test urban designs under varying conditions, ensuring that pedestrian-friendly elements remain effective across multiple contexts.

By integrating real-time simulations, VR enables planners to explore how different lighting conditions affect perceived safety, how adverse weather impacts pedestrian movement, and how high foot traffic alters navigation patterns. For example, a pedestrian pathway that appears functional under ideal conditions may become difficult to navigate when crowded or when visibility is low. VR simulations help identify these challenges early in the design process, allowing for proactive adjustments to enhance pedestrian accessibility and safety.

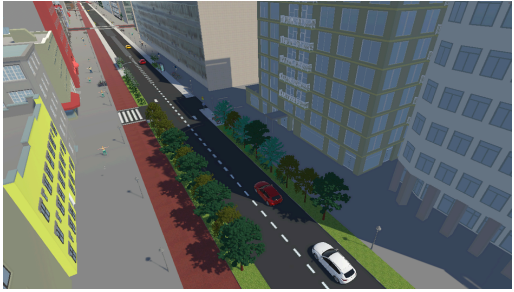

Figure 2: Scenario Simulation in VR.

Additionally, VR supports iterative design testing by enabling planners to toggle between multiple urban layouts. This capability allows designers to compare different sidewalk widths, street layouts, and public space configurations in real-time, facilitating data-driven decision-making. With this level of flexibility, VR helps optimize walkability by refining urban designs based on real user interactions rather than static theoretical models.

### 3.3 Enhanced Stakeholder Communication

Urban planning involves multiple stakeholders, including policymakers, designers, developers, and the general public. However, traditional planning methods often rely on complex architectural drawings and technical reports, making it difficult for non-experts to fully understand proposed urban changes. This creates barriers to public engagement and limits the ability of communities to provide meaningful feedback.

VR enhances stakeholder communication by offering an intuitive, immersive visualization of urban designs. Instead of interpreting abstract blueprints, stakeholders can "walk" through virtual environments and experience proposed developments firsthand. This level of accessibility fosters more inclusive urban planning by enabling residents to provide feedback based on lived experience rather than technical interpretation.

For policymakers, VR serves as a valuable tool for decision-making by providing a shared visual reference that all stakeholders can understand. By facilitating real-time discussions, VR reduces miscommunication and accelerates consensus-building. Additionally, the ability to present side-by-side comparisons of different design alternatives allows for more effective negotiation, ensuring that final urban plans align with both technical feasibility and community needs.

### 3.4 Conclusion

VR's unique affordances—first-person spatial interaction, real-time scenario testing, and enhanced stakeholder engagement—position it as a transformative tool for walkability assessment in urban planning. By bridging the gap between technical analysis and experiential evaluation, VR provides a more holistic approach to designing pedestrian-friendly environments. Future advancements should focus on increasing environmental realism, integrating real-time data from Digital Twin models, and enhancing user accessibility to ensure wider adoption. By leveraging these innovations, VR has the potential to redefine urban walkability assessments and support the development of more inclusive, sustainable, and pedestrian-friendly cities.

## 4 Current and Future Research

With a focus on using VR to improve walkability in urban area, I have developed and conducted a study with a VR prototype. The VR prototype developed in this study simulates a newly renovated urban area to assess walkability through immersive, first-person interactions. Created using Unity and the Virtual Reality Toolkit (VRTK), the prototype was deployed on a Meta Quest 3 headset, allowing users to navigate a 650 × 850 m virtual environment featuring pathways, green spaces, pedestrian crossings, public seating, etc. Nine professionals from urban planning, design, and landscape architecture participated in a structured three-phase study, including pre-interaction interviews, VR engagement, and post-interaction feedback. This prototype provided an immersive environment for assessing walkability by allowing participants to navigate virtual urban spaces, experience pedestrian pathways, and identify walkability challenges in real-time. The feedback collected from these professionals has informed key areas for improvement, including enhancing environmental realism, refining user interactions, and incorporating more dynamic urban conditions. Future research will build upon the insights gained from the development and testing of the VR prototype.

The next phase of research will focus on refining the VR prototype by integrating more sophisticated environmental elements such as dynamic weather conditions, varying crowd densities, and real-time lighting changes. These enhancements will improve the accuracy of walkability assessments, allowing planners to evaluate pedestrian infrastructure under diverse conditions. Data collection within the VR environment will be expanded to include movement tracking, eye gaze analysis, and interaction logs, providing deeper insights into how planners perceive accessibility, wayfinding, and spatial organization.

Building on this, generative design methods will be incorporated into the VR system to allow for the rapid iteration of urban layouts. This approach will enable planners and stakeholders to explore multiple design alternatives, compare their impact on pedestrian movement, and refine urban plans based on experiential feedback. The iterative testing process will ensure that urban designs prioritize walkability, safety, and accessibility while considering stakeholder preferences. By allowing users to interact with AI-generated urban models, this research will further bridge the gap between data-driven urban planning and human-centered design.

Future iterations will also integrate Digital Twin technology to create a continuously evolving urban model that updates based on real-time data. This addition will enhance the predictive capabilities of the VR prototype, allowing planners to simulate and test urban interventions in response to changing pedestrian patterns, infrastructure modifications, and environmental factors. The combination of VR and Digital Twin technologies will support a more adaptive and data-informed approach to walkability assessment, ensuring that urban planning decisions align with real-world conditions. Throughout these research iterations, accessibility and usability improvements will be prioritized to ensure that VR-based urban planning tools are practical for a broad range of stakeholders. Efforts will focus on optimizing the VR interface, reducing hardware constraints, and developing cost-effective solutions to increase adoption. Ethical considerations will also be addressed by ensuring that urban simulations accurately represent diverse populations and do not reinforce biases in design representation. By iteratively refining the VR prototype and integrating advanced urban planning technologies, this research will contribute to a more inclusive, data-driven framework for pedestrian-centric urban development.

Expanding the accessibility and scalability of VR applications remains a critical challenge. High hardware costs and technical barriers hinder widespread adoption in urban planning workflows. Future research should explore cost-effective solutions, such as web-based VR platforms and lightweight, portable headsets, to make immersive urban simulations more accessible to planners, policymakers, and community members. Simplifying user interfaces and interaction methods will further lower technical barriers, ensuring that diverse stakeholders can effectively engage with VR-based urban assessments.

Ethical considerations must also be addressed to ensure that VR-based urban planning is inclusive and representative of diverse populations. Digital urban models should account for varying mobility needs, ensuring that walkability assessments consider accessibility for individuals with disabilities. Additionally, VR-generated urban proposals must be critically examined to prevent socio-economic biases that could reinforce urban inequalities. Future research should establish ethical frameworks for VR development, prioritizing fairness, inclusivity, and transparency in urban decision-making.

Advancing VR for walkability assessment requires addressing these technical and methodological challenges. Enhancing realism, integrating quantitative analytics, improving accessibility, and ensuring ethical considerations will help VR evolve into a robust decision-support tool for urban planning. By refining these aspects, VR can foster more pedestrian-friendly, inclusive, and data-driven urban development strategies.

## 5 Conclusion

This research highlights VR's potential as a transformative tool in urban walkability assessment, offering planners an immersive and interactive platform to evaluate pedestrian experiences. By providing a first-person perspective, VR bridges the gap between abstract planning methods and real-world usability, enabling more informed decision-making in urban design. The findings underscore VR's ability to enhance stakeholder engagement, facilitate iterative feedback, and reveal walkability issues that traditional planning tools often overlook.

Despite its advantages, further advancements are necessary to improve realism, usability, and scalability. Integrating dynamic environmental conditions, real-time pedestrian interactions, and data-driven generative design can enhance VR's effectiveness in simulating urban environments. Additionally, increasing accessibility through cost-effective solutions and refining user interfaces will ensure broader adoption across planning disciplines.

As urbanization accelerates, VR presents an opportunity to redefine pedestrian-centered design by incorporating experiential insights into planning processes. Its integration with Digital Twin and Generative Design can create adaptive, data-driven urban models that respond to evolving urban challenges. Continued research should focus on refining VR's capabilities, ensuring its role as a standard tool in sustainable, inclusive, and walkable city development.

## 6 Acknowledgments

This project has received funding from the European Union's Horizon 2020 research and innovation programme under the Marie Skłodowska-Curie grant agreement No 101034328.